\documentclass{article}
\usepackage{amsmath,amssymb}
\usepackage{graphicx}
\usepackage{url}

\begin{document}

\title{Cryptanalysis of the SASI Ultralightweight RFID Authentication Protocol with Modular Rotations}

\author{Julio Cesar Hernandez-Castro$^\dagger$\, Juan M. E.
Tapiador$^\dagger$, \\Pedro Peris-Lopez $^\dagger$ and
Jean-Jacques Quisquater$^\ddagger$ \\
\\
$^\ddagger$  Crypto Group, DICE,Universite Louvain-la-Neuve Place du
Levant,\\ 1 B-1348 Louvain-la-Neuve, Belgium \\
$^\dagger$ Computer Science Department, Carlos III University\\
Avda. de la Universidad, 30, 28911 Leganes, Madrid, Spain}


\maketitle

\begin{abstract}
In this work we present the first passive attack over the SASI lightweight authentication protocol with modular rotations. This can be used to fully recover the secret $ID$ of the RFID tag, which is the value the protocol is designed to conceal. The attack is described initially for recovering $\lfloor log_2(96) \rfloor=6$ bits of the secret value $ID$, a result that by itself allows to mount traceability attacks on any given tag. However, the proposed scheme can be extended to obtain any amount of bits of the secret $ID$, provided a sufficiently large number of successful consecutive sessions are eavesdropped. We also present results on the attack's efficiency, and some ideas to secure this version of the SASI protocol.\\


\end{abstract}

\section{Introduction}

In 2007 Hung-Yu Chien published a very interesting ultralightweight
authentication protocol providing Strong Authentication and Strong
Integrity (SASI) for very low-cost RFID tags \cite{Chien}.

This was a much needed answer to the increasing need for schemes
providing such properties in very constrained environments like RFID
systems. As the previous attempts to design ultralightweight
protocols have failed (all proposals have been broken), this new
scheme was specially interesting.

As we will see later, the major difference between this proposal and existing ones is the inclusion of the rotation
operation. There has been, however, some confusion over the concrete type of rotation recommended by the author. It is important to note that the way in which rotations should be performed is not specified at all in the original paper \cite{Chien}. So the first researchers to publish some weaknesses (two desynchronization attacks) against the protocol \cite{Sun} needed to contact the author to clarify this issue. After a private communication, the author stated that the rotation he intended to use in the protocol was $Rot(A,B)= A << wt(B)$, where $wt(B)$ stands for the Hamming weight of vector $B$.

That turned out to be a wise decision, as if he had decided to use the more common rotation definition of $Rot(A,B)= A << B$ mod $N$, he would have run into the attack described in this paper. This latter version of the protocol, with a modular rotation instead of a hamming weight rotation, is the one which is analyzed in this work.

The rest of the paper is organized as follows. In the next section we describe the SASI protocol, then in Section \ref{Sec:MainResults} we introduce our attack. Finally, in Section \ref{Sec:Conclusions} we extract some conclusions that could help in devising new and stronger versions of this variant of the SASI protocol. The source code of a very simple implementation of the attack can be found in the Appendix.

\section{Description of the SASI Protocol}
The SASI protocol is briefly described in the following, where $R$
represents the reader, $T$ represents the tag, $IDS$ stands for an
index pseudonym, $ID$ is tag's private identifier, $K_i$ represent
tag's secret keys and $n_1$ and $n_2$ are nonces. The $ID$ is the
most valuable information allowing the unequivocally identification
of tagged items, a property that is not provided by other consolidated
identification systems such as barcodes.

\begin{enumerate}
  \item  $R \rightarrow T : hello$\\
  \item  $T \rightarrow R : IDS$\\
  \item  With $IDS$, the reader finds in the backend database the tag's secret values $ID$, $K_1$, and $K_2$.\\
  \item $R$ generates nonces $n_1$ and $n_2$ to construct messages $A$, $B$ and $C$ as follows\\

    $A=IDS \oplus K_1 \oplus n_1$\\
    $B=(IDS \vee K_2) + n_2$\\
    $C=(K_1 \oplus \bar{K_2})+(K_2 \oplus \bar{K_1})$, where\\
        $\bar{K_1}=Rot(K_1 \oplus n_2, K_1)$\\
        $\bar{K_2}=Rot(K_2 \oplus n_1, K_2)$\\

where $\oplus$ stands for the usual addition modulo 2, $+$ represents addition modulo $2^{96}$, and $\vee$ is the usual bitwise \textbf{or} operation.

Finally, the reader sends to the tag the concatenation of $A$, $B$ and $C$\\
$R \rightarrow T: A\|B\|C$\\

  \item From $A$ and $B$, respectively, the tag can obtain values $n_1$ and $n_2$. Then, it locally computes $C$ and checks if the result of its local computation is equal to the sent value. If this were the case, it updates the values of $IDS$, $K_1$ and $K_2$ in the following manner:\\

    $IDS^{next}=(IDS+ID)\oplus(n_2\oplus\bar{K_1})$\\
    $K_1^{next}=\bar{K_1}$\\ $K_2^{next}=\bar{K_2}$\\

  \item $T \rightarrow R : D$ with\\

    $D=(\bar{K_2}+ID)\oplus((K_1\oplus K_2)\vee\bar{K_1})$\\

  \item $R$ verifies $D$ and, if it is equal to the result of its local computation, it updates $IDS, K_1$ and $K_2$ just as the tag.\\
\end{enumerate}

\section{Cryptanalysis of SASI with Modular Rotations}\label{Sec:MainResults}

Before presenting the cryptanalysis of SASI with modular rotations, we explain the background and general assumptions in which the protocol is based.

\subsection{Background}
In 2006, Peris et al. proposed a family of Ultralightweight Mutual
Authentication Protocols (henceforth referred to as the UMAP family
of protocols). Chronologically, M$^{2}$AP \cite{PerisHER-2006-uic}
was the first proposal, followed by EMAP
\cite{PerisHER-2006-rfidsec} and LMAP \cite{PerisHER-2006-otm-is}.
Although some vulnerabilities were discovered (active attacks
\cite{LiW-2007-sec, Chiencypta-2007} and later on passive attacks
\cite{Barasz-RFIDSEC2007,Barasz-EURASIP2007}) which rendered those first proposals insecure, they were
an interesting advance in the field of lightweight cryptography for low-cost RFID tags.

The SASI protocol is highly reminiscent of the UMAP family, and more
concretely, of the LMAP protocol.



Before the SASI proposal, however, all the messages exchanged over the insecure radio channel were computed by the composition of very simple operations such as addition modulo 2, addition modulo $2^{96}$, and bitwise operations like OR and AND. This presented a major drawback, as all of these operations are triangular functions (T-functions) \cite{Klimov}. That is, these functions have the property that output bits only depend of the leftmost input bits, instead of all input bits. Furthermore, the composition of triangular operations always results in a triangular function. This undesirable characteristic greatly facilitated the analysis of the messages transmitted in the UMAP family of protocols, and thus the work of the cryptanalyst.

The main difference between LMAP and SASI is the inclusion of a non-triangular function, such that the composition of all operations would no longer be triangular. Specifically, rotation is now included in the set of operations supported by the tag, which is reasonable assumption, as it can be performed quite efficiently. 


 \subsection{Analytical Results}



The natural way of attacking this protocol is to consider what
happens when modular rotations are not performed, that is, when the amount
of rotation given by the second argument is zero modulo 96. For
these cases, the proposed protocol uses exactly the same set of
operations that lead to the attacks over the previous ultra
lightweight protocols, that is, no triangular functions. This should ease any analysis. 
Therefore:
\begin{equation}
\begin{array}{ll}
\bar{K_1} & = Rot(K_1 \oplus n_2, K_1) = Rot(K_1 \oplus n_2, K_1 \mathrm{~mod~} 96)\\
          & = Rot(K_1 \oplus n_2, 0) = K_1 \oplus n_2\\
\end{array}
\end{equation}

  Similarly,

\begin{equation}
\bar{K_2}=Rot(K_2 \oplus n_1, K_2)=K_2 \oplus n_1\\
\end{equation}

This has a particularly nasty impact in the process of index pseudonym ($IDS$) update, since

\begin{equation}
\begin{array}{ll}
    IDS^{next} & =(IDS+ID)\oplus(n_2\oplus\bar{K_1})\\
               & =(IDS+ID)\oplus(n_2 \oplus\ K_1 \oplus n_2)\\
               & =(IDS+ID)\oplus {K_1}\\
\end{array}
\end{equation}

    So we have that $ID=IDS^{next}\oplus {K_1}-IDS$ and we can take full advantage of the knowledge that $K_1=K_2=0$ mod $96$ to conclude that, with a probability depicted in Table 1, only depending on the value of $N$ ($N=96$ in this case, but other values could be used for recovering more bits) it holds that

\begin{equation}
ID \mathrm{~mod~} 96 \approx (IDS^{next}-IDS) \mathrm{~mod~} 96
\end{equation}

    As both values $IDS^{next}$ and $IDS$ are public and easily observable by snooping at two consecutive authentication sessions, this relation allows us to recover the $\lfloor log_2(96) \rfloor = 6$ less significant bits of the secret $ID$ and, analogously, to perform a traceability attack over the .\\

    The only question that remains is how to recognize when the conditions $K_1=0 \mathrm{~mod~} 96$ and $K_2=0 \mathrm{~mod~} 96$ hold simultaneously, since $K_1$ and $K_2$ are secrets that only the tag and the reader should know. Fortunately, this is possible by checking if certain relations (that only involve public values) hold.\\

    Let us suppose that $K_1=K_2=0 \mathrm{~mod~} 96$ then
\begin{eqnarray}
        \bar{K_1}=Rot(K_1 \oplus n_2, K_1)=Rot(K_1 \oplus n_2, 0)=K_1 \oplus n_2\\
        \bar{K_2}=Rot(K_2 \oplus n_1, K_2)=Rot(K_2 \oplus n_1, 0)=K_2 \oplus n_1
\end{eqnarray}
    So
\begin{eqnarray}
\begin{array}{ll}
    C & =(K_1 \oplus \bar{K_2})+(K_2 \oplus \bar{K_1}) \\
    & =K_1 \oplus K_2 \oplus n_1 + K_2 \oplus K_1 \oplus n_2
\end{array}
\end{eqnarray}
    which implies that
\begin{equation}
\begin{array}{ll}
C \mathrm{~mod~} 96 & =K_1 \oplus K_2 \oplus n_1 + K_2 \oplus K_1
\oplus n_2 \mathrm{~mod~} 96 \\&  \approx n_1 + n_2 \mathrm{~mod~}
96. \end{array}
\end{equation}
    The value of $n_1 + n_2 \mathrm{~mod~} 96$ can also be probabilistically obtained from the observed values of public messages $A$, $B$ and $IDS$ because:
\begin{equation}
    A=IDS \oplus K_1 \oplus n_1 \Rightarrow n_1=A \oplus IDS \oplus K_1
\end{equation}
and then we can get that
\begin{equation}
\begin{array}{ll}
n_1 \mathrm{~mod~} 96 & = A \oplus IDS \oplus K_1 \mathrm{~mod~} 96
\\& \approx A \oplus IDS \mathrm{~mod~} 96 \end{array}
\end{equation}
because, by hypothesis, $K_1=0 \mathrm{~mod~} 96$\\
    Similarly, we can obtain that, as $B=(IDS \vee K_2) + n_2$,  then
\begin{equation}
n_2 \approx (B-IDS) \mathrm{~mod~} 96
\end{equation}

    All in all, we can conclude that if $K_1=K_2=0 \mathrm{~mod~} 96$ then, with a probability given in Table 1
\begin{equation}\label{eq1}
\begin{array}{ll}
    C \mathrm{~mod~} 96 & \approx n_1 + n_2 \mathrm{~mod~} 96 \\ & \approx (A \oplus IDS)+ (B-IDS) \mathrm{~mod~} 96
\end{array}
\end{equation}
    so what is only left is to passively snoop multiple authentication sessions and, for each one, verify if the above condition holds. If this is the case, one should compute the value $(IDS^{next}-IDS) \mathrm{~mod~} 96$ and from this, approximate $ID \mathrm{~mod~} 96$. \\

\begin{table}
\caption{Probabilities of Equations 4, 8, 10, 11 and 12
simultaneously holding for different values of $N$, given that
$K_1=K_2=0 \mathrm{~mod~} N$}
\begin{center}
\begin{tabular}{|c|c|c|c|c|}
\hline
  N & $2^t$ & $3 \cdot 2^t$ & $4 \cdot t+10$ & $2 \cdot t+5$\\
  \hline
  Probability & $1.00$ & $0.33$ & $2 \cdot N^{-1}$ & $N^{-1}$ \\
    \hline
\end{tabular}
\label{Tprob}
\end{center}
\end{table}

Only one last tweak is needed to perform a successful attack: Just
by chance, the above relation will be true even if the two
preconditions $K_1=0 \mathrm{~mod~} 96$ and $K_2=0 \mathrm{~mod~}
96$ are not simultaneously true, and this will lead us to a possibly
wrong estimation for $ID \mathrm{~mod~} 96$.

This is, however, easily fixable by simply observing many values of
$(IDS_{next}-IDS) \mathrm{~mod~} 96$ when equation (\ref{eq1})
holds, because the true value of $ID \mathrm{~mod~} 96$ will likely
be the most common.

This fact has been experimentally verified and leads to the attack
schematically described in Fig. 2.

\begin{figure*}
\begin{center}
\begin{tabular}{l}
\hline
~1.~~For $i=0$ to $96$\\
~2.~~~~~~~$Observations[i]=0$\\
~3.~~Repeat a sufficiently high number of times $N$ the following steps:\\
~4.~~~~~~~Observe an authentication session and get $IDS$, $A$, $B$ and $C$\\
~5.~~~~~~~Check if for these values it holds that $C = (A \oplus IDS)+ (B-IDS) \mathrm{~mod~} 96$\\
~6.~~~~~~~If this is not the case, go to step 4.\\
~7.~~~~~~~Perform the following tasks:\\
~8.~~~~~~~~~~~~~~Wait for the authentication session to finish.\\
~9.~~~~~~~~~~~~~~Send the tag a hello message to obtain $IDS^{next}$.\\
10.~~~~~~~~~~~~~~Compute $c=(IDS^{next}-IDS) \mathrm{~mod~} 96$\\
11.~~~~~~~~~~~~~~Increment $Observations[c]$\\
12.~~Find $m$, the maximum of the values in $Observations[i]$.\\
13.~~Conjecture that $m = ID \mathrm{~mod~} 96$.\\
\hline
\end{tabular}
\end{center}
\begin{center}
Fig. 2. Outline of the attack. \end{center} \label{Fig1}
\end{figure*}

\subsection{Efficiency analysis}

The attack presented could be performed not only for recovering $\lfloor log_2(96) \rfloor$ bits of the secret value $ID$, but also works for other modulus, with varying probabilities as shown in Table 1. In particular, the set of probabilistic equations (i.e. equations 4, 8, 10, 11, 12) all hold with probability one for modulus that are a power of 2, so this allows for more efficient attacks able of obtaining much more bits (i.e. $log_2(256)=8$, $log_2(512)=9$, $log_2(1024)=10$, etc.) if needed. In these cases, we naturally need to observe more authentication sessions for recovering more $ID$ bits.\\

As a rule of thumb we have concluded, after extensive experimentation, that an attacker following this procedure is on average able of recovering the $\lfloor log_2(S)\rfloor$ least significant bits of $ID$ after observing around $\theta(S)$ authentication sessions.

\section{Concluding Remarks} \label{Sec:Conclusions}

In this article we have presented an attack against a variant of a novel and quite
interesting ultralightweight authentication protocol.

We analyze the SASI protocol under the assumption that the most common rotation definition (i.e. modular rotation) is employed. This analysis points out that the inclusion of the rotation operation (a non-triangular function) is a necessary but by itself not sufficient condition to achieve security in lightweight protocols. It also highlights the advantages of the hamming rotation over the modular rotation here explored, namely that the former is much less likely to behave like the identity. This could be a good reason to lead future designers of ultralightweight protocols towards a preference for the hamming over the modular rotation.

We have to acknowledge, however, that the proposed attack is not successful against the hamming rotation as advocated by the author of the protocol. To day, authors do not know any other passive attack against the SASI protocol or its modular variant. Active attacks, on the other hand, abound both against the hamming and against the modular version of the protocol. First, Sun et al. proposed to desynchronization attacks. Then, in \cite{Cao} it was proposaed a denial-of-service and traceability attack. Recently, D'Arco et al. proposed another desynchronization attack \cite{Darco}, an identity disclosure attack, and finally a full disclosure attack against modular SASI.

Some different design decisions would, on the other hand, have considerably harden our attack, and we will briefly
describe then in the following:
\begin{itemize}
  \item The $IDS$ updating could be improved as it is dependant of $n_2$ and $\bar{K_1}$ which is again a function of $n_2$. This is instrumental in our attack and, in any case, leads to all sorts of bad statistical properties.
  \item The definitions of $\bar{K_1}$  and $\bar{K_2}$ should be rethought, as in the current way there is a kind of distributive property ($\bar{K_1}=Rot(K_1 \oplus n_2, K_1)=Rot(K_1,K_1) \oplus Rot(n_2, K_1)$ ) that could ease attacks.

This can be avoided by, for example, using addition instead of xor as the inner operator, although part of the problem still remains. The ideal solution should be to devise a more complex key scheduling, but of course this will have an additional cost in terms of gate equivalents and performance.

  \item The use of the bitwise OR operation should be performed with extreme care, as the resulting messages are strongly biased.

As an example, in the current protocol definition $n_2$ could be approximated simply by computing $n_2 \approx B-\mathbf{1}$.

Message $D$ suffers from a similar problem. The use of a bitwise AND operation would produce similar undesirable
effects. Past experience with other lightweight protocols has shown that these two operators should only be included in the inner parts of the algorithm, and every effort should be made to disguise their output into seemingly random output when constructing public messages such as $B$ and $D$.
\end{itemize}

In fact, an even more general version of this attack is possible.  This alternative is, on the other hand,
significantly less efficient than the attack scheme described here. It consists simply in observing and storing the different values of equation 4 (regular rotations is assumed again). In a well-designed protocol, these should approximately follow a uniform distribution, but we have experimentally observed that this is far from being the case. Following this extremely simple approach, with no approximations nor preconditions, we are able to recover up to 4 bits of the secret $ID$ after around $2^{10}$ authentication sessions with a 100\% success probability, a fact that could lead to a very straightforward tracking attack.\\

Finally, we can conclude that the SASI protocol is indeed an interesting step in the right direction towards fully secure ultralightweight protocols, and that the decision about what type of rotations to employ was a correct one because if modular rotations were used instead, the resulting protocol will fall short of the security requirements typically needed in these schemes.\\

\appendix{\textbf{Appendix A: Attack's source code}}

This is the source code of our attack, implemented in Python

\begin{verbatim}

#Traceability & recovery attack against the Modular SASI 
#Ultralightweight Authentication Protocol

from random import *
from scipy import *

NumExperiments=2**18

def wt(a):
    w=0
    while a:
        if a%2: w=w+1
        a=a>>1
    return w

def rot(a,b):
    return ((((a << b) % 2**96) | (a >> (96-b)) % 2**96)) % 2**96

def sasiprotocol(L):
    IDS, SID, N1, N2, K1, K2 = L[0], L[1], L[2], L[3], L[4], L[5]
    A=IDS^K1^N1
    B=((IDS | K2)+ N2) % 2**96
    K1hat=rot(K1^N2, K1%96)
    K2hat=rot(K2^N1, K2%96)
    C=((K1^K2hat)+(K2^K1hat)) % 2**96
    D=((K2hat+SID)% 2**96)^((K1^K2)|K1hat)
    IDSnext=((IDS+SID)%2**96)^(N2^K1hat)
    O = [A%2**96, B%2**96, C%2**96, D%2**96, IDSnext%2**96, K1hat%2**96, K2hat%2**96]
    return O

#The secret value we will try to obtain is I[1]=SID

I=[]
for i in range(6):
	I.append(randint(0,(2**96)-1))

#Keep the value of I for the future, so copy it on nI and only manipulate wI
wI=I

Observations=[]
for i in range(96):
	Observations.append(0)
j=0
for i in range(NumExperiments):
	O=sasiprotocol(wI)
	#Get IDS
	IDS=wI[0]
	#Get A, B, C
	A=O[0]
	B=O[1]
	C=O[2]
	#Check if it holds that C=(A^IDS)+(B-IDS)%96
	if (C%96==((A^IDS)+(B-IDS))%96):
		j=j+1
		#Obtain the value of IDSnext
		IDSnext=O[4]
		#Compute c=(IDSnext-IDS)%96
		c=(IDSnext-IDS)%96
		Observations[c]=Observations[c]+1

	#Then, a new protocol session begins
	wI=[O[4],wI[1],randint(0,(2**96)-1),randint(0,(2**96)-1),O[5],O[6]]

#Print Observations & Compute the maximum
max=0
for i in range(96):
	print "The value ", i, "has been observed ", Observations[i], "times"
	if (Observations[i]>max):
		max=Observations[i]
		maxindex=i
print "The probability of a useful session is, approx.=1/",NumExperiments/j*(1.0)
print "The maximum value, and our guess for SID%96 is ", maxindex
print "The correct value of SID%96 is ", I[1]%96
print "The difference between this values is", abs(I[1]%96-maxindex)	
#This difference is always a power of two meaning that 
#the least significant bits of our guess were correct
\end{verbatim}


\begin{thebibliography}{99}


\bibitem{Chien}
Hung-Yu Chien. ``SASI: A New Ultralightweight RFID Authentication
Protocol Providing Strong Authentication and Strong Integrity''.
\textit{IEEE Transactions on Dependable and Secure Computing}
4(4):337--340. Oct.-Dec. 2007.

\bibitem{PerisHER-2006-uic}
P.~Peris-Lopez, J. C. Hernandez-Castro, J. M. Estevez-Tapiador, and
A.~Ribagorda.
\newblock {M2AP}: A minimalist mutual-authentication protocol for low-cost
  {RFID} tags.
\newblock In {\em Proc. of UIC'06}, volume 4159 of {\em LNCS}, pages
912--923. Springer-Verlag, 2006.

\bibitem{PerisHER-2006-rfidsec}
P.~Peris-Lopez, J. C. Hernandez-Castro, J. M. Estevez-Tapiador, and
A.~Ribagorda.
\newblock {LMAP}: A real lightweight mutual authentication protocol for
  low-cost {RFID} tags.
\newblock Hand. of Workshop on RFID and Lightweight Crypto, 2006.

\bibitem{PerisHER-2006-otm-is}
P.~Peris-Lopez, J. C. Hernandez-Castro, J. M. Estevez-Tapiador, and
  A.~Ribagorda.
\newblock {EMAP}: An efficient mutual authentication protocol for low-cost
  {RFID} tags.
\newblock In {\em Proc. of IS'06}, volume 4277 of {\em LNCS}, pages
352--361. Springer-Verlag, 2006.

\bibitem{LiW-2007-sec}
T.~Li and G.~Wang.
\newblock Security analysis of two ultra-lightweight {RFID} authentication
  protocols.
\newblock In {\em Proc. of IFIP-SEC'07}, 2007.

\bibitem{Chiencypta-2007}
C.~Hung-Yu and H.~Chen-Wei.
\newblock Security of ultra-lightweight {RFID} authentication protocols and its
  improvements.
\newblock {\em SIGOPS Oper. Syst. Rev.}, 41(4):83--86, 2007.

\bibitem{Barasz-RFIDSEC2007}
M. B\'ar\'asz,  B.  Boros, P. Ligeti, K. L\'oja, and D. Nagy.
``Breaking LMAP'', \textit{Proc. of RFIDSec'07}, 2007.


\bibitem{Barasz-EURASIP2007}
M. B\'ar\'asz,  B.  Boros, P. Ligeti, K. L\'oja, and D. Nagy.
``Passive Attack Against the {M2AP} Mutual Authentication Protocol
for {RFID} Tags'', \textit{Proc. of First International EURASIP
Workshop on RFID Technology}, 2007.

\bibitem{Klimov}
A. Klimov and A. Shamir. ``New Applications of T-functions in Block
Ciphers and Hash Functions''. \textit{Proc. of FSE'05}, LNCS vol.
3557, pp. 18--31. Springer-Verlag, 2005.


\bibitem{Sun}
Hung-Min Sun, Wei-Chih Ting, and King-Hang Wang. ``On the Security
of Chien's Ultralightweight RFID Authentication Protocol''.
Cryptology ePrint Archive. \url{http://eprint.iacr.org/2008/083},
2008.

\bibitem{Cao}
Tianjie Cao, Elisa Bertino, and Hong Lei. ``Security Analysis of the
SASI Protocol''. \textit{IEEE Transactions on Dependable and Secure
Computing}, 2008.

\bibitem{Darco}
Paolo D'Arco and Alfredo De Santis. ``From Weaknesses to Secret
Disclosure in a Recent Ultra-Lightweight  RFID Authentication
Protocol''. Cryptology ePrint Archive.
\url{http://eprint.iacr.org/2008/470}, 2008.


\end{thebibliography}
\end{document}